\documentstyle[color,twocolumn,prl,aps,epsfig,amssymb]{revtex}

\definecolor{Red}{cmyk}{0,1,1,0}

\definecolor{Blue}{cmyk}{1,1,0,0}

\newcommand{\lt}{\left}
\newcommand{\rt}{\right}

\newcommand{\nn}{\nonumber\\}

\newcommand{\eq}[1]{(\ref{#1})}

\newcommand{\mev}{\mbox{MeV}}
\newcommand{\gev}{\mbox{GeV}}
\newcommand{\ov}[1]{\overline{#1}}

\newcommand{\be}{\begin{equation}}
\newcommand{\ee}{\end{equation}}
\newcommand{\br}{\begin{eqnarray}}
\newcommand{\er}{\end{eqnarray}}
\newcommand{\ba}{\begin{array}}
\newcommand{\ea}{\end{array}}
\newcommand{\bi}{\begin{itemize}}
\newcommand{\ei}{\end{itemize}}
\newcommand{\bn}{\begin{enumerate}}
\newcommand{\en}{\end{enumerate}}
\newcommand{\bc}{\begin{center}}
\newcommand{\ec}{\end{center}}

\newcommand{\bs}{$B_s\rightarrow\mu^+\mu^-$}
\newcommand{\cbs}{{\cal B}(B_s\rightarrow\mu^+\mu^-)}
\tighten

\newcommand{\gsim}{\lower.7ex\hbox{$\;\stackrel{\textstyle>}{\sim}\;$}}
\newcommand{\lsim}{\lower.7ex\hbox{$\;\stackrel{\textstyle<}{\sim}\;$}}

\begin{document}
\tolerance=100000
\thispagestyle{empty}



\twocolumn[\hsize\textwidth\columnwidth\hsize\csname@twocolumnfalse\endcsname

%
\title{\vspace{-3ex}{ \small CERN--TH/2001-211 
                      \hfill hep-ph/0108037 }\\[2mm]
Correlation of $B_s \to \mu^+\mu^-$ and $(g-2)_\mu$ in Minimal
  Supergravity}
\author{Athanasios Dedes$^1$, Herbert K.~Dreiner$^1$ and 
 Ulrich Nierste$^2$}
\address{ 
 {\it $^1$ Physikalisches Institut, Universit\"at Bonn, Nu{\ss}allee 12, 
D-53115 Bonn, Germany} \\
 {\it $^2$ CERN, TH Division, CH-1211 Geneva 23, Switzerland} 
} 
\maketitle
\begin{abstract}
  We analyse the rare decay mode \bs\ in the minimal supergravity
  scenario (mSUGRA). We find a strong correlation with the muon
  anomalous magnetic moment $(g-2)_\mu$. An interpretation of the
  recently measured excess in $(g-2)_\mu$ in terms of mSUGRA
  corrections implies a substantial supersymmetric enhancement of the
  branching ratio $\cbs$: if $(g-2)_\mu$ exceeds the Standard Model
  prediction by $4\cdot 10^{-9}$, 
  $\cbs$ is larger by a factor of { 10--100}
  than in the Standard Model and within reach of Run-II of the
  Tevatron. Thus an experimental search for \bs\ is a stringent test of
  the mSUGRA GUT scale boundary conditions. If the decay \bs\ is
  observed at Run-II of the Tevatron, then we predict the mass of the
  lightest supersymmetric Higgs boson to be less than 120 GeV. The
  decay \bs\ can also significantly probe the favoured parameter range
  in SO(10) SUSY GUT models.
\end{abstract}
]

Supersymmetry (SUSY) is an attractive and widely studied extension of
the Standard Model (SM).  The minimal supergravity model (mSUGRA)
\cite{mSUGRA} relates all supersymmetric parameters to just 5 real
quantities: the universal scalar and gaugino masses $M_0$ and
$M_{1/2}$, the trilinear term $A_0$, the ratio $\tan \beta$ of the two
Higgs vacuum expectation values, and $\mbox{sgn}\,\mu$, where $\mu$ is
the Higgs\-ino mass parameter. The first three 
 quantities are defined at a high,
grand unified energy scale 
and the others at the electroweak scale. 
They are the boundary conditions for the
renormalization group equations, which determine the physical
parameters at our low scale. Precision observables, which are affected
by SUSY corrections through loop effects, play an important role in
constraining the supersymmetric parameter space. The small number of
parameters makes mSUGRA highly predictive so it can be significantly
tested by low energy precision measurements. In this letter we show
that the decay \bs\ is a stringent test of the mSUGRA scenario, in
particular when correlated with $(g-2)_\mu$.

Recently the Brookhaven National Laboratory (BNL) reported an excess
of the muon anomalous magnetic moment $a_{\mu}=(g-2)_{\mu}/2$ over its
SM value \cite{BNL}. The difference $\delta a_{\mu} = a_{\mu}^{exp} -
a_{\mu}^{SM} = (43\pm 16) \cdot 10^{-10}$ corresponds to a 2.6$\sigma$
deviation from the SM. An mSUGRA interpretation of this anomaly
implies $\mu >0$ (in the sign convention with $M_{1/2}>0$ and equal
signs of the diagonal elements of the chargino mass matrix)
\cite{Lopez:1994vi}. It further invites a large $\tan\beta\gsim10$
\cite{Feng}. The discrepancy in the case of $a_\mu$ is by itself not
significant enough to justify the claim of new physics, especially
since the calculation of $a_{\mu}^{SM}$ involves two hadronic
quantities: the hadronic contributions to the photon self-energy,
which must be obtained from other experiments, and the (smaller)
light-by-light scattering contribution, which can only be estimated
with hadronic models. A more conservative estimate of the latter would
reduce the BNL anomaly to a 2$\sigma$ effect \cite{BNL}.  Hence in
order to resolve the possible ambiguity between mSUGRA and alternative
explanations of $\delta a_{\mu}$ one ideally wishes to study other
observables whose sensitivity to supersymmetric loop corrections is
correlated with $\delta a_{\mu}$. It is our purpose here to show the
strong correlation between $\cbs$ and $\delta a_{ \mu}$ in mSUGRA.

SUSY modifies B meson observables if $\tan \beta$ is large, because
the $b$ Yukawa coupling becomes sizable.~Especially sensitive are
quantities with a $b$ quark chir\-ality flip like the branching ratios
${\cal B}(B \to X_s \gamma)$ and ${\cal B} (B \to \ell^+ \ell^-)$. In
mSUGRA the low energy value for the trilinear term $A_t$ is dominated
by $M_{1/2}$ with $A_t <0$ for $M_{1/2}>0$ \cite{Masiero}. Then
$\mu>0$ implies that the charged-Higgs-top loop and the chargino-stop
loop tend to cancel in ${\cal B}(B \to X_s \gamma)$, so that the
sensitivity to mSUGRA corrections is weakened.  A further disadvantage
of this decay mode is that it requires an experimental cut on the
photon energy, which introduces some hadronic uncertainty.

In \cite{Masiero} the possible impact of flavour-blind SUSY on other B
physics observables, in particular those which enter the fit of the
unitarity triangle, were studied and only small effects were found.
This did not include the decay \bs.  In contrast to the observables in
\cite{Masiero}, the branching ratio $\cbs$ grows like $\tan^6\beta$
\cite{Babu,Chankowski,Urban}, with a possible several orders of
magnitude enhancement. We here go beyond this work to study \bs\ in the
mSUGRA model. Since ${\cal B} (B \to \ell^+ \ell^-) \propto
m_{\ell}^2$, the branching ratio is largest for $\ell=\tau$.  Yet
$\tau$'s are hard to detect at hadron colliders, so that the prime
experimental focus is on the search for \bs. B factories running on
the $\Upsilon(4S)$ resonance produce no $B_s$ mesons. Leptonic
branching ratios of $B_d$ mesons are smaller by a factor of $|V_{td}
/V_{ts}|^2\lsim 0.06$.  Since in B factories the boost of the $B_d$
meson is known and the considered leptonic decay rates can be
substantially enhanced over their SM values in SUSY, we encourage our
colleagues at BaBar and BELLE to look for $B_d \to \tau^+ \tau^-$
decays, as well. From now on we restrict ourselves to the decay mode
\bs.

In \cite{Babu,Chankowski,Urban} the SUSY corrections to $\cbs$ were
calculated at the one-loop level. For large $\tan \beta$, higher order
corrections can be large, eventually of order 1. In \cite{cgnw} $\tan
\beta$-enhanced supersymmetric QCD corrections have been summed to all
orders in perturbation theory. 
We have incorporated these dominant higher order corrections
 by replacing the $b$ Yukawa coupling
$h_b\propto m_b \tan \beta$ with $ h_b^{{\rm eff}} = h_b/(1+\Delta
m_b)$, where $\Delta m_b \propto \mu \tan \beta$ depends on the gluino
and sbottom masses and can be found in \cite{cgnw}.  $\Delta m_b$ is
positive for $\mu>0$.  The dominant contribution to $\cbs $ is
proportional to $h_b^{{\rm eff}\,4}$, so that the inclusion of $\Delta
m_b$ tempers the large-$\tan \beta$ behaviour.

The considered branching ratio can be expressed as
\begin{eqnarray}
\lefteqn{\!\! \cbs \; =} \nn
&& 6.0 \cdot 10^{-7} 
\lt(\frac{|V_{ts}|}{0.040}\rt)^2
\lt(\frac{f_{B_s}}{230\,\mev}\rt)^2
\frac{m_{\mu}^2}{m_{B_s}^2}
\sqrt{1-\frac{4 m_\mu^2}{m_{B_s}^2}} \nonumber \\
 && \Biggl \{
 \biggl ( 1-\frac{4 m_\mu^2}{M_{B_s}^2}\biggr ) 
  \lt| \frac{m_{B_s}^2 C_S}{m_{\mu}} \rt|^2 + 
 \lt| \frac{m_{B_s}^2 C_P}{m_{\mu}} - 2 C_A \rt|^2 \Biggr \} .
 \label{brancing}
\end{eqnarray}
Here $|V_{ts}|=0.040\pm 0.002$ is the relevant CKM matrix element and
$f_{B_s}=(230 \pm 30)\; \mev$ \cite{fbs} is the $B_s$ decay constant.
In \eq{brancing} we have kept the dependence on the lepton mass
$m_{\mu}$, so that the generalisation to $B_d \to \tau^+ \tau^-$ is
straightforward. The Wilson coefficients $C_S$, $C_P$ and $C_A$, which
contain the short-distance physics, are normalised as in
\cite{Nierste}.  The coefficients $c_S$, $c_P$ and $c_{10}$ defined in
\cite{Urban} are related to ours by $C_S=-2 c_S \sin^2 \theta_W$, $C_A
= -2 c_{10} \sin^2 \theta_W$ and $C_P = 2 c_P \sin^2 \theta_W$. Within
the SM, $C_S$ and $C_P$ are negligibly small and the NLO result for
$C_A$ can be well approximated by $C_A=2.01 (\ov{m}_t/167\;
\gev)^{1.55}$ \cite{Buchalla}. Here $\ov{m}_t\equiv \ov{m}_t(m_t) $ is
the top quark mass in the $\ov{\rm MS}$ scheme.  $\ov{m}_t = 167$
GeV corresponds to a pole mass of $m_t=175\; \gev$.  The SM prediction
is given by $\cbs = (3.7 \pm 1.2) \cdot 10^ {-9}$, with the
uncertainty ($\pm 25\%$) dominated by $f_{B_s}$.  This is also the
main hadronic uncertainty in the SUSY calculation.

During Run-I of the Tevatron, CDF determined  \cite{CDFbmumu}
\begin{eqnarray}
\cbs < 2.6 \times 10^{-6},
\qquad \mbox{ at 95\% C.L.} 
\label{expbmumu}
\end{eqnarray}
The single event sensitivity of CDF at Run-IIa is estimated to be $1.0
\cdot 10^{-8}$, for an integrated luminosity of 2 fb$^{-1}$
\cite{fnalbrep}. Thus if mSUGRA corrections enhance $\cbs$ to
e.g.\ $5\cdot 10^{-7}$, one will see 50 events in Run-IIa. Run-IIb may
collect 10-20 fb$^{-1}$ of integrated luminosity, which implies 
250-500 events in this example.

In SUSY, the dominant coefficients are $C_{S,P}$ since they are
proportional to $\tan^3\beta$.~We would like to understand the effect
of the restricted mSUGRA parameters on $C_{S,P}$ and thus on $\cbs$.
In mSUGRA the low-energy values of both $\mu$ and the squark masses
are dominated by the (GUT scale) value of $M_{1/2}$ through the
renormalization group equations.  For not-too-large $M_0,M_{1/2}\lsim
500$ GeV and $A_0\simeq 0$ GeV we can derive the approximate formula
$\cbs \approx 10^{-6} \tan^6 \beta \, M_{1/2}^2 \gev^4 /(M_{1/2}^2 +
M_0^2)^3$. In the vicinity of the maximum (near $M_{1/2}=0.4\, M_0$)
the approximate formula is not accurate.  A similar estimate of the
supersymmetric contribution to $a_\mu$ yields $(\delta
a_\mu)_{SUSY}\propto\tan\beta f(M_0)/M_{1/2} ^2$.  $(\delta
a_\mu)_{SUSY}$ depends on slepton masses, which are less sensitive to
$M_{1/2}$ than squark masses; they are dominated by $M_0$. We have
encoded the $M_0$ dependence in the slowly varying function $f(M_0)$.
{ Hence both $\cbs$ and $(\delta a_\mu)_{SUSY}$ grow with $\tan \beta$
  and decrease with increasing $M_{1/2}$.  For this} it is essential
that we have made the assumption of the mSUGRA GUT scale boundary
conditions.  Thus within mSUGRA we expect a strong correlation between
$\cbs$ and $(\delta a_\mu)_{SUSY}$: if $\delta a_\mu \not=0$ requires
a SUSY explanation with large $\tan\beta$ then we would expect $\cbs$
to be strongly enhanced. If, however, the supersymmetric explanation
of $\delta a_\mu$ requires small $M_{1/2}$ and a moderate value of
$\tan\beta$ then we would expect only a moderate enhancement of
$\cbs$.
\begin{figure}[t]
\centerline{\psfig{figure=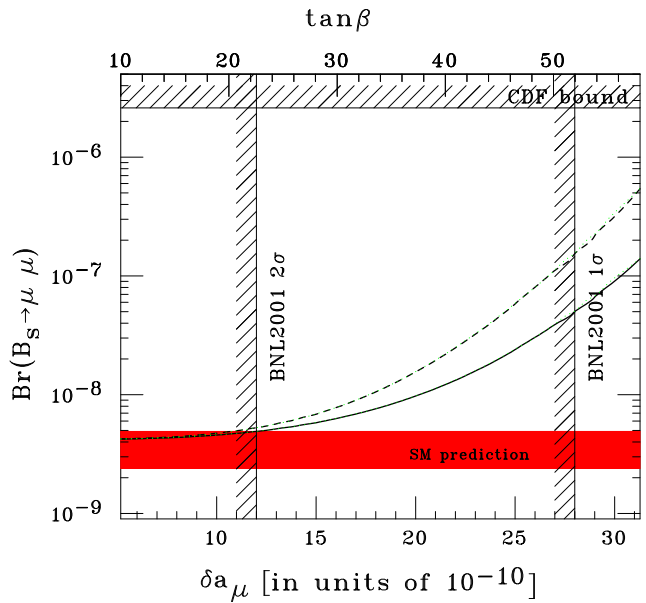, angle=90,height=8cm}}
\caption{$(\delta a_\mu)_{SUSY}$, versus $\cbs$ for 
  $\tan\beta$ (top) and $M_{1/2}$=450,
  $M_0=350, A_0=0, \mu>0, m_t=175$ GeV.  Shown also, the SM
  prediction, the present bound by CDF~\protect\cite{CDFbmumu}, on
  $\cbs$ as well as the present 1$\sigma$ and 2$\sigma$ bound on
  $\delta a_\mu$ from BNL~\protect\cite{BNL}. We used $f_{B_s}=230$
  MeV.}
\label{fig1}
\end{figure}

We now study these effects quantitatively.~For this we use the full
computation of Eq.(\ref{brancing}) including the resummed SUSY QCD
corrections, and restricting ourselves to the mSUGRA parameters.~In
Fig.\ref{fig1}, we show the direct correlation between $\cbs$ and
$(\delta a_\mu)_{SUSY}$ for the fixed parameters: $M_{1/2}=450\; \gev,
M_0=350\; \gev, A_0=0, \mu>0$ and $m_t=175$ GeV.  On the upper edge we
show the $\tan\beta$ dependence. We restrict ourselves to
$\tan\beta<58$ in order to guarantee radiative electroweak symmetry
breaking (REWSB). We have included the SM prediction and the CDF bound
from Eq.(\ref{expbmumu}). The solid (dashed) curve represents the
$\cbs$ result with (without) resummation of the $\tan \beta$-enhanced
SUSY-QCD corrections.  In this example, the resummation suppresses
$\cbs$ by 75\% for $\tan\beta\gsim 50$. In order for mSUGRA to account
for $\delta a_\mu$ within 1$\sigma$ of the current BNL measurement at
this parameter point, we see that we need a large value of $\tan\beta
\gsim 50$. Due to the strong correlation within mSUGRA we then predict
$\cbs \gsim 5 \cdot 10^{-8}$, which is observable by CDF at Run~II.

\begin{figure}[t]
\centerline{\psfig{figure=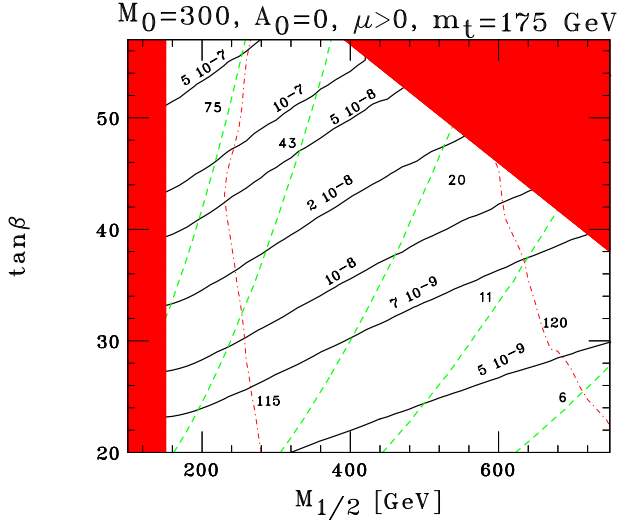, angle=90,height=7cm}}
\caption{Contours of $\cbs$ (solid) and $(\delta a_\mu)_{SUSY}$ 
  (in units $10^{-10})$ (dashed) in the $M_{1/2}$-$\tan\beta$ plane.
  The lightest neutral CP-even Higgs mass is shown as well
  (dot-dashed). The shaded regions are excluded, as described in the
  text. The mSUGRA parameters are given at the top.}
\label{fig2}
\end{figure}

\begin{figure}
\centerline{\psfig{figure=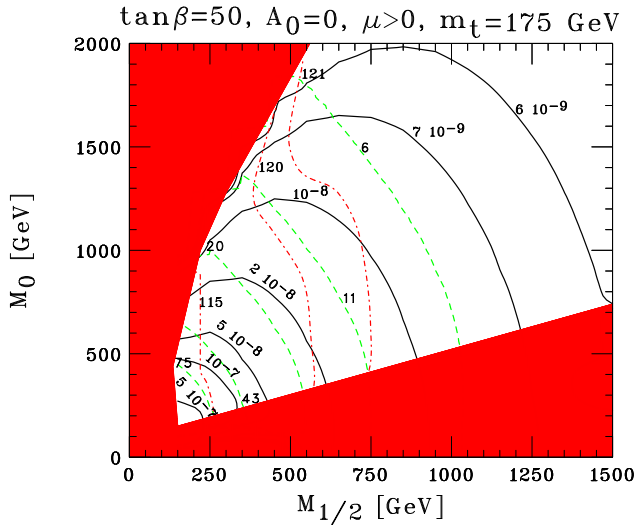, angle=90,height=7cm}}
\caption{Contour plots  of the $\cbs$ (solid) and on $(\delta a_
  \mu)_{SUSY}$ (dashed)  in the ($M_{0},M_{1/2}$)-plane for
  mSUGRA parameter values  as shown. The shaded regions are excluded as
  described in the text. Contours of the light Higgs boson mass
  (dot-dash line) are also shown. }
\label{fig3}
\end{figure}

As we discussed above, we expect $\cbs$ to dominantly depend on the
mSUGRA parameters $M_{1/2}$ and $\tan\beta$. In Fig.\ref{fig2} we show
the $\cbs$ (solid) and the $(\delta a_\mu)_{SUSY}$ (dashed) contours
in this plane. We have fixed: $M_0=300$ GeV, $A_0=0$, $\mu>0$ and
$m_t=175$ GeV. The $2\sigma$ contours for $\delta a_\mu$ (11,75) are
explicitly given. The left vertical shaded region is theoretically
excluded since it does not allow for REWSB or violates the LEP
chargino bound.  The upper right triangular shaded region is excluded,
since the LSP is not neutral.  If as expected, CDF can probe down to
$\cbs\gsim2\cdot 10^{-7}$ at RUN~IIa, this corresponds to a
sensitivity of $(M_{1/2}, \tan\beta)$ ranging from $(160\,\gev, 47)$
to $(450\,\gev, 57)$. The qualitative discussion of before is now
nicely reproduced.  $\cbs$ decreases with increasing $M_{1/2}$ and
rapidly increases with $\tan\beta$. Fig.\ref{fig2} also nicely shows
the cross-correlation between $(\delta a_\mu)_{SUSY}$ and $\cbs$. If
both $\cbs$ and $(\delta a_\mu)_{SUSY}$ are found in disagreement with
the SM and are measured with a $50\%$ and $20\%$ accuracy,
respectively, then for given $M_0$, this fixes $\tan \beta$ to better
than $ 20\%$ and $M_{1/2}$ to better than $30\%$.

It is conventional to discuss mSUGRA physics in the $(M_{1/2},M_0
)$-plane. In Fig.\ref{fig3} we show the contours of $\cbs$ (solid) and
$(\delta a_\mu)_{SUSY}$ (dashed) in this plane, for $\tan\beta=50$,
$A_0=0$, $\mu>0$ and $m_t=175 \, \gev$. Again we include the CDF bound
Eq.(\ref{expbmumu}) and the Higgs mass contours.  The left shaded
region is excluded through the requirement of REWSB or the chargino
bound.  The lower right shaded region is excluded through the
requirement of a neutral LSP. A sensitivity of $\cbs\gsim
2\cdot10^{-7}$ at CDF now corresponds to a sensitivity of
$M_{1/2}\lsim 280 \,\gev$ and $M_0\lsim 400 { \,\gev}$, respectively.

While CDF is not able to see squark masses directly up to 0.7 TeV
(corresponding to $M_{1/2}=M_0\simeq 300$ GeV,), it will nevertheless
be able to prepare the ground for LHC by observing the \bs\ mode.
Even better, after 10$\,{\rm fb}^{-1}$ CDF will probe $M_{1/2} \lsim
450$ GeV and $M_0 \lsim 600\,\gev$ (for $\tan \beta=50$) which in
mSUGRA corresponds to masses for the heaviest superpartners of 1 TeV.
We conclude the discussion of Fig.\ref{fig3} with the prediction of
the light Higgs boson mass $M_h$ (dot-dashed line) for $\tan\beta=50$
in the mSUGRA scenario~\cite{ASBS}.  Any measurement of $\cbs$ by
itself implies a useful \emph{upper}\ bound on $M_h$. The simultaneous
information of $\cbs$ and $\delta a_{ \mu}$ fixes $M_h$ in most
regions of the $(M_{1/2},M_0)$-plane.  A Higgs mass around 115.6 GeV
results in $ 10^{-8} \lsim \cbs \lsim 3\cdot 10^{-7}$ which would most
likely be measured before the Higgs boson is discovered.

In Figs.~1-3 we have chosen $A_0=0$. A non-zero $A_0$ changes the
value of $A_t$ at low energies. This parameter plays a crucial role
for the GIM cancellations among the contributions of different squarks
to $\cbs$. Changing $A_0$ to $-500\,\gev$ in the scenario of Fig.~1
enhances $\cbs$ by up to a factor of 6 compared to the case with
$A_0=0$. For $A_0=+500$ GeV $\cbs$ is slightly decreased.

In our figures we have omitted further constraints on the mSUGRA
parameter space, in order to clearly show the correlation between
$\cbs$ and $(\delta a_\mu)_{SUSY}$. The most significant further
constraint comes from the measurement of ${\cal B}(B\rightarrow
X_s\gamma)$ \cite{bsgammaexp}, whose prediction is less certain in the
large $\tan\beta$ region \cite{bsgammath,Masiero}.~If we take the
conservative approach of \cite{Djouadi:2001yk}, then we can exclude
values of $M_{1/2}\lsim 250\,\gev$ in Fig.\ref{fig2} for $\tan\beta
\gsim 25$.  In the scenario of Fig.\ref{fig3} this implies $\cbs
  \lsim 5\cdot10^{-7}$.  For a discussion of the constraints from
supersymmetric dark matter see for example \cite{Feng,Djouadi:2001yk}
and references therein.

The large values of $\tan\beta$ we have been considering are
theoretically well motivated within SUSY SO(10) Yukawa
unification.~There a narrow parameter region can explain the observed
$\delta a_\mu$ while still being consistent with the constraint from
$b\rightarrow s\gamma$ \cite{Baer,Stuart}.~This is {\it not} within
the context of mSUGRA. However, in this parameter region both $\mu$
and $M_{1/2}$ are light, while the CP-odd Higgs boson mass is less
than 300 GeV, and $\tan\beta\approx 50$. Therefore we expect $\cbs$ to
be strongly enhanced.~As an example we determine $\cbs$ for the best
fit points found in~\cite{Stuart}: $M_A=110$ GeV, $m_{\tilde{\chi}_1}
\lsim 250$ GeV, $|A_t| \gsim 1$ TeV, $m_{\tilde{t}} \lsim 1$ TeV and
$\tan\beta \simeq 50$. Within the hadronic uncertainties $\cbs \gsim
10^{-5}$ which is already excluded by CDF~\cite{CDFbmumu}. Thus the
SO(10) models should be reconsidered in the light of $\cbs$.  Turning
it around, if an SO(10) GUT model is the correct description of nature
then the decay \bs\ must be just around the corner.
 
In conclusion, we have found a striking correlation between the muon
anomalous magnetic moment $a_{\mu}$ and the branching ratio $\cbs$ in
mSUGRA scenarios. If the reported excess in $a_{\mu}$ \cite{BNL} is
caused by mSUGRA corrections with large $\tan\beta$, one faces more
than an order of magnitude enhancement of $\cbs$ over its SM value.
This is within reach of Run-II of the Tevatron. The combined
measurements significantly constrain the mSUGRA parameters, allowing a
determination of $\tan\beta$ and $M_{1/2}$. A measurement of $\cbs$
will further constrain the mass of the lightest Higgs bosons. SO(10)
SUSY explanations of the measured $a_{\mu}$ are barely compatible with
the present upper bound on $\cbs$.

{\it We thank
   B. Dutta, G. Isidori, K. Mizukoshi and X. Ta\-ta for cross-checking our
  numerical results.
  A.D. acknowledges financial support from the Network
  RTN European Program HPRN-CT-2000-00148. }


\newpage

\end{document}